\documentclass[5p,times,twocolumn]{elsarticle} 

\usepackage{amssymb}
\usepackage{geometry}                		
\usepackage{graphicx}				
\usepackage{subfigure}                          
\usepackage{url}

\begin{document}

\begin{frontmatter}


\title{Thermal neutron flux measurements in STAR experimental hall.}
\author[BNL]{Y. Fisyak}
\author[UCLA]{O. Tsai}
\author[BNL]{Z. Xu}

\address[BNL]{Brookhaven National Laboratory, Upton, New York 11973}
\address[UCLA]{University of California - Los Angeles, Physics Department, Los Angeles, CA 90095-1547}

\begin{abstract}
We report on measurements of thermal neutron fluxes at different locations 
in the STAR experimental hall during  pp $\sqrt s$ = 510~GeV Run 13 at RHIC.
We compared these measurements with calculations based on \textsc{PYTHIA} as minimum bias events generator, 
the detailed GEANT3 simulation of the STAR detector and the experimental hall, and using GCALOR as neutron transport code.  
A good (within $\approx$ 30\%) agreement was found at locations near ($\approx$1m) and very far ($\approx$10m) from the beam pipe. 
For intermediate locations ($\approx$ 5m) the simulation overestimates neutron flux by a factor of $\approx$3.
\end{abstract}


\end{frontmatter}

\section{Introduction} 
\par
Since the time of $R\&D$ for the SSC detectors \cite{Groom,Diwan} it has been understood that the main source of background in a detector at modern colliders are
collisions at interaction point. The contribution from other sources (beam gas interactions, beam halo particles, etc.) estimated 
to be below 10\%\cite{Mokhov}.
Extensive simulations of background conditions were part of detectors optimizations for SSC and LHC experiments. 
ATLAS\cite{ATLAS} and CMS\cite{CMS}  have made simulations for all types of backgrounds including neutrons. 
Estimations of the neutron fluxes in experimental areas were based on simulations only, 
without support from experimentally measured data. 
Only recently  the ATLAS-MPX collaboration\cite{ATLAS-MPX} published results of absolute background  measurements 
in the ATLAS experimental hall
including thermal neutrons and 
made a comparison with results of simulations with GEANT3+GCALOR\cite{GCALOR} and Fluka\cite{FLUKA}. Their conclusion was \cite{ATLAS-MPX} :
``Measured thermal neutron fluxes are found to be largely in agreement with the original simulations, mostly within a factor of two. Significant deviations are observed in the low radiation regions of ATLAS cavern, where measured thermal neutron fluxes are found to be lower than predicted by Monte Carlo simulations.''
\par
The STAR detector at the Relativistic Heavy Ion Collider (RHIC)\cite{STAR} is planning series of upgrades in the near future with detectors using different types of silicon sensors. 
Reliable estimations of neutron background at STAR are required to evaluate different technologies for these upgrades. 
This necessity and the lack of experimental results for neutron background estimates were our motivations for this work.
Same questions have been raised in context of ongoing detector R\&D for proposed Electron Ion Collider (EIC\cite{EIC}):
\begin{itemize}
  \setlength{\itemsep}{0cm}
  \setlength{\parskip}{0cm}
\item 
What are neutron background conditions currently at the STAR detector and will be at EIC?
\item
How reliable can we estimate these conditions ?
\end{itemize}
To answer these questions we:
\begin{itemize}
  \setlength{\itemsep}{0cm}
  \setlength{\parskip}{0cm}
\item
made measurement of the absolute thermal neutron flux at different locations in the STAR\cite{STAR}  
Wide Angle Hall (WAH) during RHIC Run 13\cite{RunXIII},
\item 
compared experimental results with simulation in order to understand how reliable this simulation is, and
\item 
estimated fluxes of the intermediate energy neutrons using simulation results.
\end{itemize}
For the purpose of future discussions we will classify neutrons by kinetic energy($E_{kin}$) as follows: 
\begin{itemize} 
  \setlength{\itemsep}{0cm}
  \setlength{\parskip}{0cm}
\item
intermediate energy neutrons with $E_{kin}$ in range 100~keV$\textendash$1~MeV, which are most damaging for electronics and silicon detectors, and 
\item 
thermal neutrons with $E_{kin}$ below 250 meV. This definition includes cold ($<25meV$),  
thermal as such ($25 meV$), and part of epithermal (25~meV$<E_{kin}<$400~meV) neutrons. The thermal neutrons generate $\gamma-$quanta producing noise in detector elements.
\end{itemize}
\section{Measurements}
\subsection{$He^3$ counter}
\begin{figure}[ht]
   \centering
   \includegraphics[width=9cm, clip=true,trim=1.7cm 2.cm 3.2cm 2.6cm]{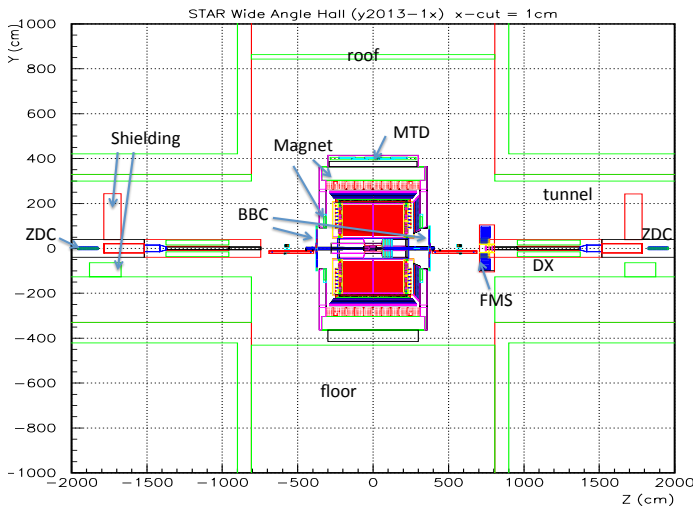} 
   \caption{STAR Wide Angle Hall GEANT3 geometry model (version y2013-1x) including building elements (floor, roof, and walls), 
     tunnel, shielding, RHIC dipole magnet (DX), and the whole STAR detector. MTD stands for Muon Telescope Detector, ZDC - Zero Degree Calorimeters, BBC - Beam Beam Counters, FMS - Forward Meson Spectrometer.}
   \label{fig:WAH}
\end{figure}

\begin{figure}[htbp]
    \centering
    \includegraphics[width=7cm , clip=true,trim=4.cm 2.cm 1.5cm 2.5cm]{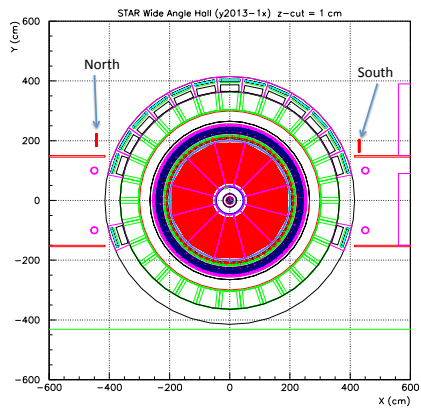} 
    \caption{\textbf{South} (x = 428 cm, y = 183 cm, z = 0) and
      \textbf{North} (x =  -442 cm, y = 202 cm, z = 0 ) locations of $He^{3}C$.}
    \label{fig:SouthNorth}
 \end{figure}

\begin{figure}[htbp]
    \centering
    \includegraphics[width=7cm, clip=true,trim=4.cm 2cm 1.5cm 2.5cm]{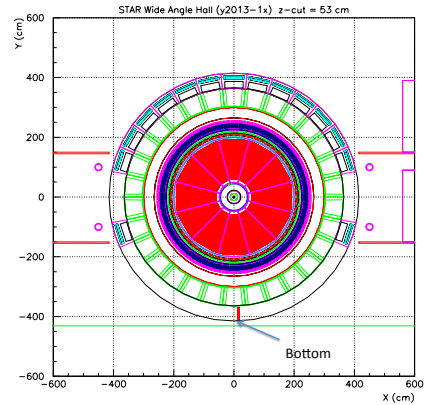} 
    \caption{\textbf{Bottom} (x = 15 cm, y = -390 cm, z = 53 cm) location  of $He^{3}C$.}
    \label{fig:Bottom}
 \end{figure}

\begin{figure}[htbp]
    \centering
    \includegraphics[width=7cm, clip=true,trim=4.cm 2cm 1.5cm 2.5cm]{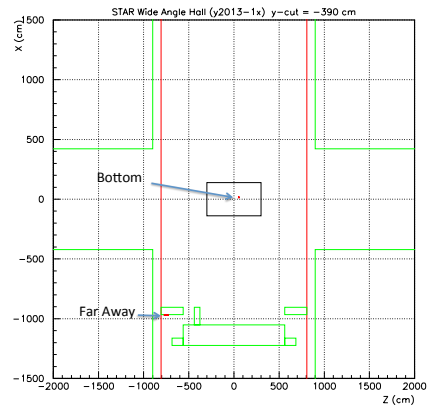} 
    \caption{\textbf{Bottom} (x = 15 cm, y = -390 cm, z = 53 cm) and \textbf{Far Away} (x = -970 cm, y = -390 cm, z = -750 cm) locations  of $He^{3}C$.}
    \label{fig:FarAway}
 \end{figure}

\begin{figure}[htbp]
   \centering
    \includegraphics[width=7cm, clip=true,trim=4.cm 2cm 1.5cm 2.5cm]{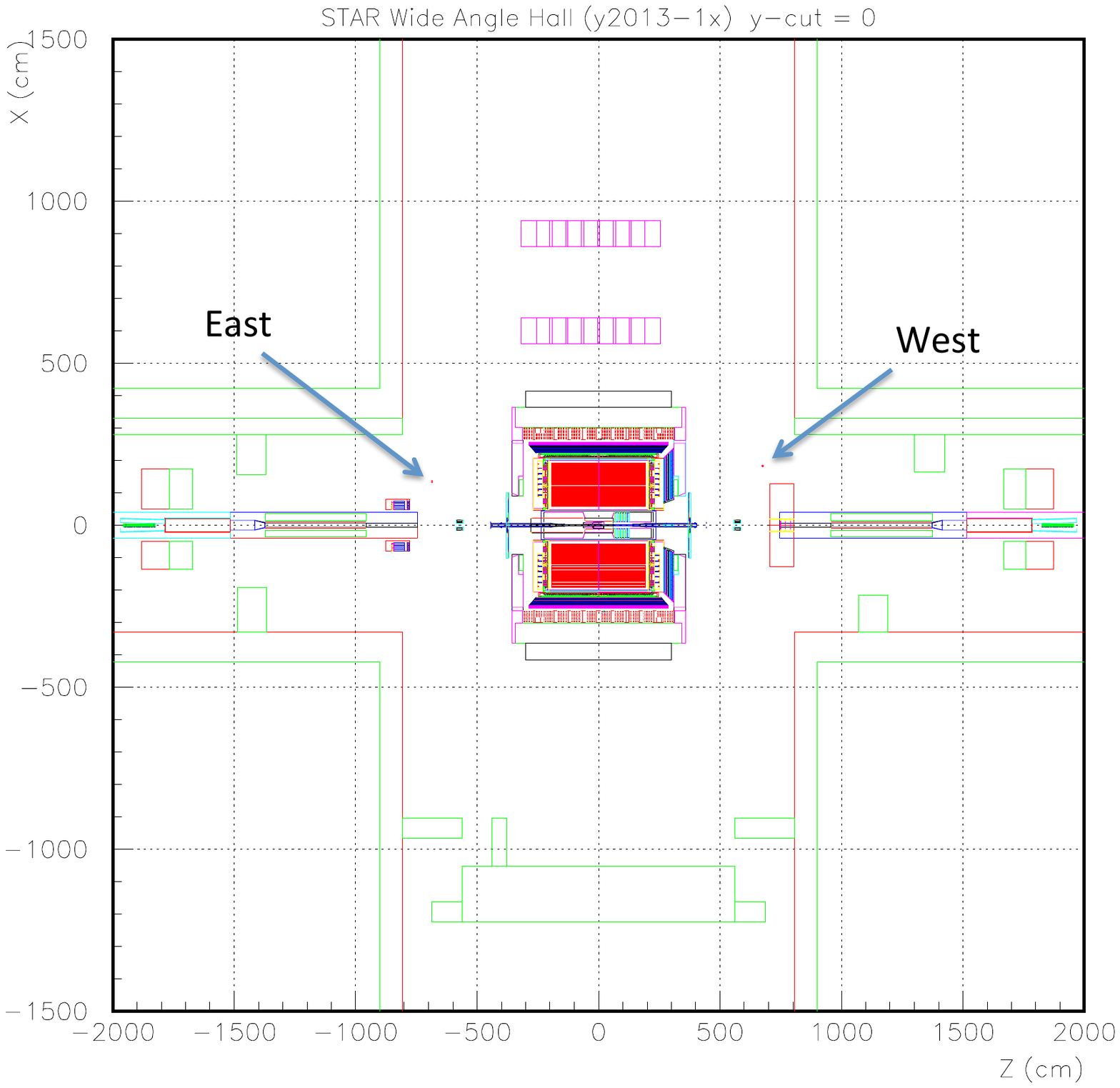} 
    \caption{\textbf{West} (x = 183 cm, y = 0, z = 676 cm) and \textbf{East} (x = 135 cm, y = -20cm, z = -686 cm)  locations of $He^{3}C$.}
    \label{fig:WestEast}
 \end{figure}

We used  a $He^3$ filled proportional counter\cite{He3} ($He^{3}C$), loaned to us by BNL Instrumentation Division, to measure fluxes of thermal neutrons in WAH.
\begin{itemize} 
 \setlength{\itemsep}{0cm}%
 \setlength{\parskip}{0cm}%
\item The thermal neutron were  detected via reaction:
$$
  n +  He^{3} \rightarrow  H^{1} +  H^{3} + 764 keV, 
$$
with cross section : $\sigma = 5.4 \sqrt(25.3~ meV/E_{kin})$ [kbarn]\cite{He3CS}.  
\item The $He^{3}C$ specification\cite{He3} gave the neutron sensitivity $100 \pm 10$ counts per $1Hz /cm^2$ of thermal neutron flux.
This sensitivity was measured with calibrated isotropic thermal neutron flux at a temperature of $25^0$C\cite{Johnson}.  
\item The signal was shaped with the threshold set to 20\% of the maximum signal (764 keV), which 
corresponds to an unambiguous thermal neutron registration (contamination of $\gamma$ and charged particles 
were due to only multiple hits during signal collection time of the detector $\approx$ 5 $\mu s$ and neglected herein).
\item During the run $He^{3}C$ was positioned at 6 locations\cite{STARcoord}  of WAH (Fig.\ref{fig:WAH}): 
the \textbf{South} and  \textbf{North}  (Fig.\ref{fig:SouthNorth}) on the level of the second platform just outside of MTD 
from the south and north sides of the detector,
the \textbf{Bottom} on the floor under MTD (Fig.\ref{fig:Bottom} and Fig.\ref{fig:FarAway}), 
the \textbf{West} and \textbf{East} near the entrances to the tunnel (Fig.\ref{fig:WestEast}),
and the \textbf{Far Away} (Fig.\ref{fig:FarAway}) on the floor just after the entrance to WAH.
\end{itemize}

\par
\begin{figure}[htbp]
   \centering
\includegraphics*[width=9cm, clip=true,trim=1.5cm 1.5cm 1.0cm 1cm]{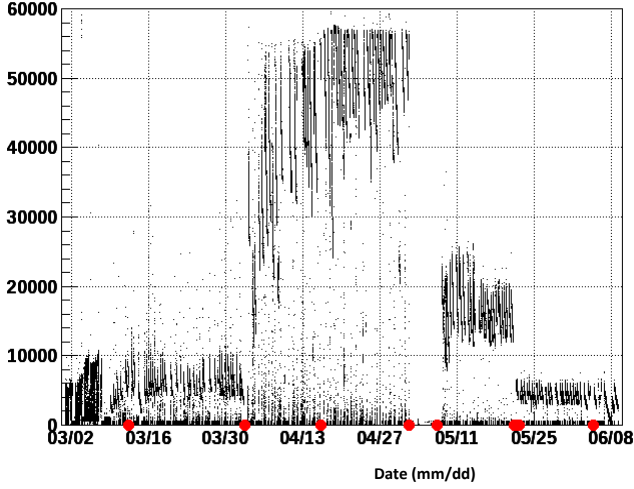} 
   \caption{Measured $He^{3}C$  rate (Hz) versus date at   different counter locations: 
\textbf{South} (during period: 03/13-04/03), 
\textbf{West} (04/03-04/17), 
\textbf{East} (04/17-05/03),   
\textbf{North}  (05/08-05/22), 
\textbf{Bottom} (05/23-06/05), and
\textbf{Far Away}  (06/06-06/10).
The location change is marked as red dots.
}   
   \label{fig:CountRate}
   \end{figure}

The shaped $He^{3}C$ signal was fed to  the so called STAR RICH scalers (channel 16),  
and the rate of the scaler (Hz) was recorded in STAR online database 
(each 15 s) and in STAR daq stream (with frequency 1 Hz) 
together with others scalers (particularly, ZDC West, ZDC East, and ZDC West and East coincidence). 
The $He^{3}C$  rate versus date of data taking for different counter locations is shown in Fig.\ref{fig:CountRate}.

\begin{figure}[htbp]
   \centering
   \includegraphics[width=9cm, clip=true,trim=0.5cm 0.0cm 0.50cm 0.25cm]{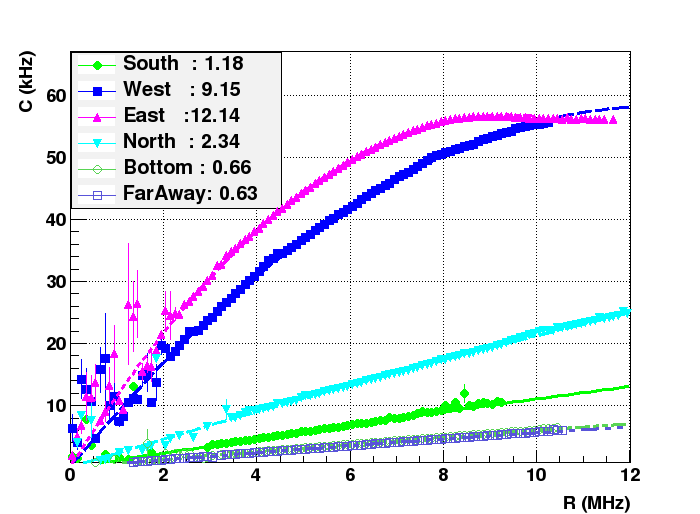} 
   \caption{The $He^{3}C$ rate (\textbf{C}) versus event rate (\textbf{$R$})  for  different counter locations. 
     The corrected counter rates (\textbf{$C_{0}$}, see text)  per 1 MHz of inelastic events at different locations are given in kHz.}
   \label{fig:Fluxes}
\end{figure}
\subsection{Event rate}
\par
In this study we used the East and West ZDC scalers.
In order to estimate event rate [MHz] the following  approach\cite{Dunlop,CDF} was used:
\begin{itemize}
 \setlength{\itemsep}{0cm}%
 \setlength{\parskip}{0cm}%
\item $N_{BC} = 9.383 \times 111/120$:  number of bunch crossings,
\item $N_{WE}$: number of crossings that contain a coincidence of the West and East counters with probability\\
 $P_{WE} = N_{WE} / N_{BC}$,
\item $N_{E}$: number of crossings that contain a hit in the East counter, $P_{E} = N_{E} / N_{BC}$,
\item  $N_{W}$: number of crossings that contain a hit in the West counter, $P_{W} = N_{W} / N_{BC}$,
\item $P_{A}$:  a probability to produce an East hit,
\item $P_{B}$:  a probability to produce a West hit,
\item $P_{AB}$: a probability to produce at least one or more  East and West coincidences in the beam crossing.
\end{itemize}
Then we used 3 equations:
\begin{eqnarray*}
P_{E} = P_{A} + P_{AB} \times (1  - P_{A} )\\
P_{W} = P_{B} + P_{AB} \times (1  - P_{B} )\\
P_{WE} = P_{A} \times P_{B} + P_{AB} \times (1 - P_{A} \times P_{B})
\end{eqnarray*}
and solved them with respect to $P_{AB}$
\begin{eqnarray*}
P_{AB} = \frac{P_{WE} - P_{E} P_{W}}{1 +  P_{WE} - P_{E} - P_{W}} = 1 - e^{-\mu} ,
\end{eqnarray*}
where $\mu$ is the mean value of Poisson distribution. 
\par
Thus  the coincidence rate (AB) corrected for random coincidence for A and B is
\begin{eqnarray*}
	N_{AB} = \mu \times  N_{BC} = -ln (1 - P_{AB}) \times N_{BC} .
\end{eqnarray*}
The coincidence rate in ZDC corresponded to $\sigma$ = 2.81 mb\cite{Dunlop} from   50 mb of pp\cite{pp510cs}  inelastic cross section at $\sqrt s$ = 510 GeV. 
Thus the total event rate: $R = 50/2.81 \times N_{AB}$.

\subsection{Fluxes}
The measured fluxes are obtained from the $He^{3}C$ rate (\textbf{C}) using the counter sensitivity. 
Dependences of the measured  \textbf{C} at the different locations on 
\textbf{$R$} 
are shown in Fig.\ref{fig:Fluxes}. 
In order to normalize \textbf{C} to 1 MHz of pp interaction rate (\textbf{$C_0$}) and also account for saturation effects in $He^{3}C$ due to its dead time, the dependences were approximated by \textbf{$C   = R \times (C_0 + R \times C_1).$}
The measurements of \textbf{$C_{0}$} for different locations are presented in Table \ref{table}.

\begin{table}[htbp]
\caption{The measured $He^{3}C$ rate (\textbf{$C_{0}$}), the estimated from the $He^{3}C$ rate neutron flux for $E_{kin} < 250 meV$ (RC) using the counter sensitivity (100$\pm$10 counts/($Hz/cm^2$)) and its efficiencies in the kinematical range ($87\%$), simulated (MC) thermal neutron flux $(Hz/cm^2)$,
 and  ratio RC to MC for the different $He^{3}C$ locations in WAH. All numbers are normalized per 1 MHz of pp inelastic collisions
at $\sqrt s$ =  510 GeV.}

\begin{center}
\begin{tabular}{|l|r|r|r|r|}
\hline
{\footnotesize  Location}&{\footnotesize \textbf{$C_{0}$} (kHz)}&{\footnotesize  RC $(Hz/cm^2)$  }&{\footnotesize   MC$(Hz/cm^2)$  }&{\footnotesize ratio }\\
\hline
{\footnotesize     South}&{\footnotesize   1.18  }&{\footnotesize   13.6 $\pm$  1.4 }&{\footnotesize    34.7 $\pm$  5.9 }&{\footnotesize   0.39 $\pm$ 0.08 }\\
{\footnotesize      West}&{\footnotesize   9.15  }&{\footnotesize  105.2 $\pm$ 10.5 }&{\footnotesize   124.1 $\pm$ 11.1 }&{\footnotesize   0.85 $\pm$ 0.11 }\\
{\footnotesize      East}&{\footnotesize  12.14  }&{\footnotesize  139.5 $\pm$ 13.9 }&{\footnotesize   105.3 $\pm$ 10.3 }&{\footnotesize   1.33 $\pm$ 0.18 }\\
{\footnotesize     North}&{\footnotesize   2.34  }&{\footnotesize   26.9 $\pm$  2.6 }&{\footnotesize    39.9 $\pm$  6.3 }&{\footnotesize   0.67 $\pm$ 0.13 }\\
{\footnotesize    Bottom}&{\footnotesize   0.66  }&{\footnotesize    7.6 $\pm$  0.8 }&{\footnotesize    23.9 $\pm$  4.9 }&{\footnotesize   0.32 $\pm$ 0.07 }\\
{\footnotesize   FarAway}&{\footnotesize   0.63  }&{\footnotesize    7.2 $\pm$  0.7 }&{\footnotesize     7.0 $\pm$  2.6 }&{\footnotesize   1.03 $\pm$ 0.40 }\\
\hline
\end{tabular}
\end{center}
\label{table}
\end{table}
\section{Simulation}
\par
To estimate fluxes,  \textsc{PYTHIA} version 6.4.26\cite{PYTHIA} as pp 510 GeV minimum biased event generator and 
GEANT3+GCALOR\cite{GCALOR} for propagation particles in WAH were used.
The STAR detector and WAH geometry description was taken as version $y2013\_1x$\cite{starsim} used for RHIC Run 13.
The only two essential changes from default STAR simulation were: 
\par
(1) reducing $E_{kin}$ cut for neutral hadrons (CUTNEU) from 1 MeV to $10^{-13}$ GeV, and
\par
(2) increasing maximum particle time of flight cut (TOFMAX) from $5\times10^{-4}$  to $1\times10^{3}$ s.
\par
\begin{figure}[htbp]
   \centering
  \includegraphics*[width=9cm, clip=true,trim=0.0cm 0.0cm 1.0cm 0.cm]{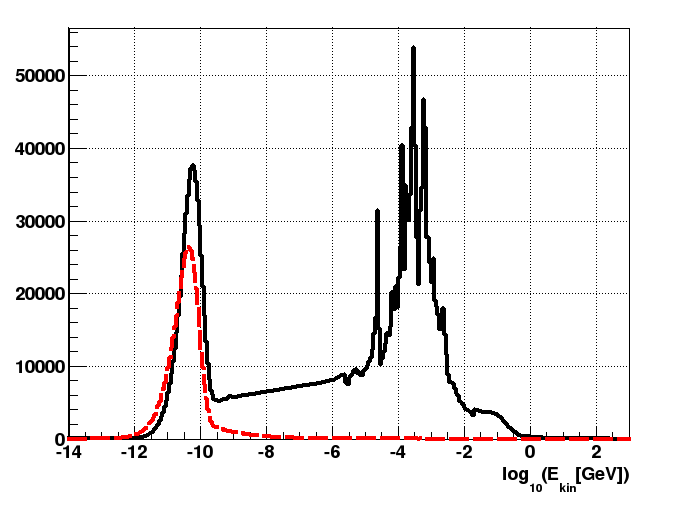} 

   \caption{The neutron  kinetic energy spectrum in WAH and result of convolution (red dashed line) of this spectrum with $He^3$ neutron cross section.
   The integral of the convoluted spectrum corresponds to $87\%$ of the total spectrum integral in region $<250meV$.}
   \label{fig:Ekin}
\end{figure}

The simulated $E_{kin}$ spectrum of neutrons in WAH and the spectrum convoluted with the neutron cross section are presented in Fig.\ref{fig:Ekin}.
From this spectrum we can conclude that $87\%$ of neutrons with  $E_{kin} < 250meV$ were detected by $He^{3}C$.
\par
\begin{figure}[htbp]
   \centering
  \includegraphics*[width=9cm, clip=true,trim=0.5cm 0.0cm 1.0cm 1.2cm]{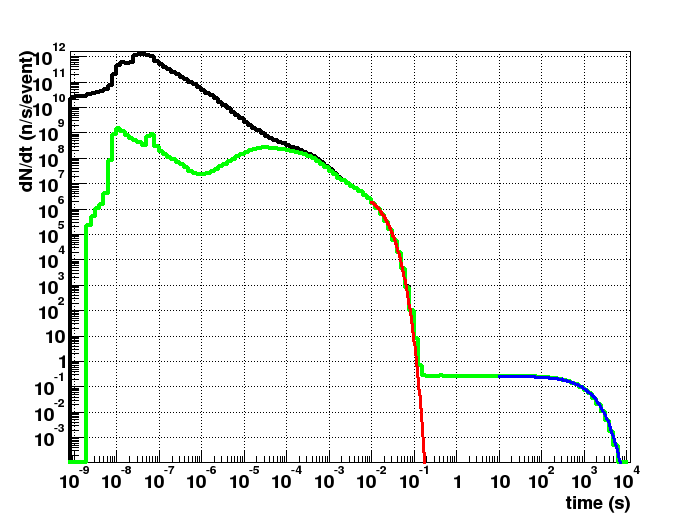} 
   \caption{The neutron's time of flight. The black and green histograms are  for all and the thermal neutrons ($E_{kin} < 250meV$), respectively.
     The red line corresponds to exponential fit $e^{-t/\tau}$ with $\tau = 7.1 ms$ and the blue line to fit with $\tau = 891 s$.}
   \label{fig:nToF}
\end{figure}

The neutron's time of flight distribution from the simulation is presented in Fig.\ref{fig:nToF}.  
There are two distinct components in the distribution: the first one with $\tau$ = 7.1 ms which corresponds to neutron dissipation from WAH 
and the second component suppressed by a factor of $10^7$ with respect to the first one  with $\tau = 891s$   
which is due to neutron decays (in the simulation  neutron life time $\tau = 887 s$ was used). 
Unfortunately, with our maximum recording rate of 1 Hz we could not detect the dissipation component. 

\par
\begin{figure}[htbp]
   \centering
   \includegraphics[keepaspectratio,scale=0.5, clip=true,trim=0.2cm 0.0cm 0.0cm 1.0cm]{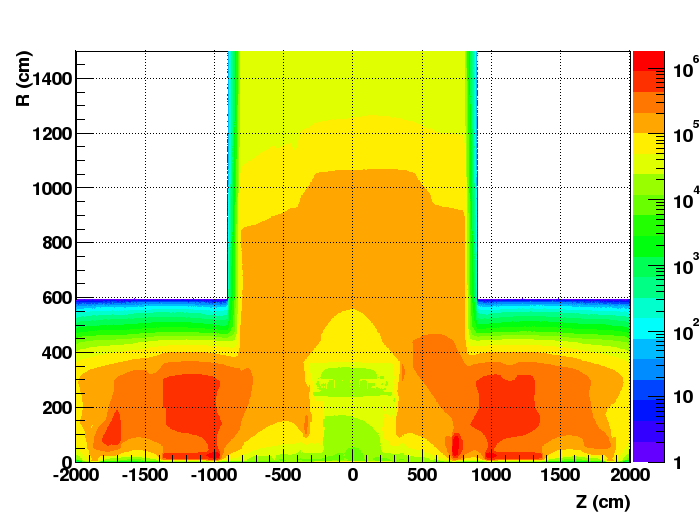} 
   \caption{Neutron flux in WAH.}
   \label{fig:Neutron}
\end{figure}
\begin{figure}[htbp]
   \centering
   \includegraphics[keepaspectratio,scale=0.5, clip=true,trim=0.2cm 0.0cm 0.0cm 1.0cm]{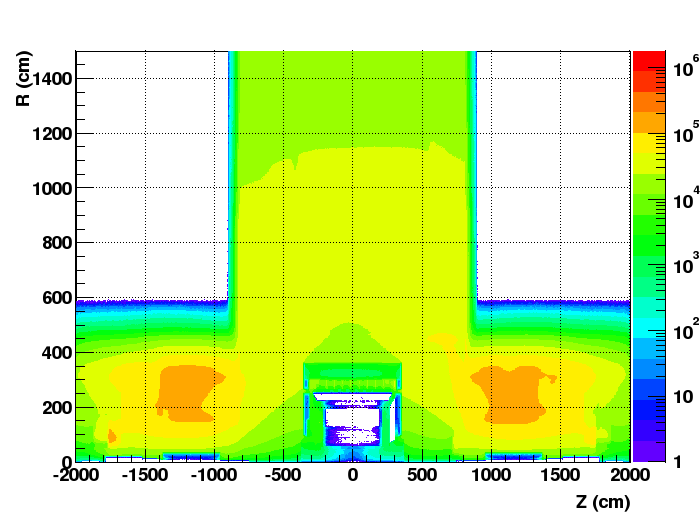} 
   \caption{Thermal neutron ($E_{kin} < 250 meV$) flux in WAH.}
   \label{fig:ThermalNeutron}
\end{figure}
\begin{figure}[ht]
  \centering
  \subfigure[]{\label{fig:NeutronsZ0}
    \includegraphics[keepaspectratio,scale=0.5, clip=true,trim=0.2cm 0.0cm 0.0cm 1.0cm]{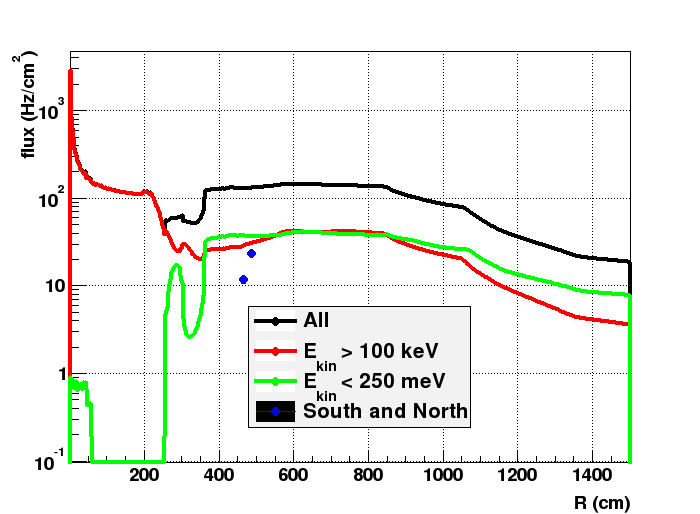}
  } 
  \hfill
  \subfigure[]{\label{fig:NeutronsZ675} 
    \includegraphics[keepaspectratio,scale=0.5, clip=true,trim=0.2cm 0.0cm 0.0cm 1.0cm]{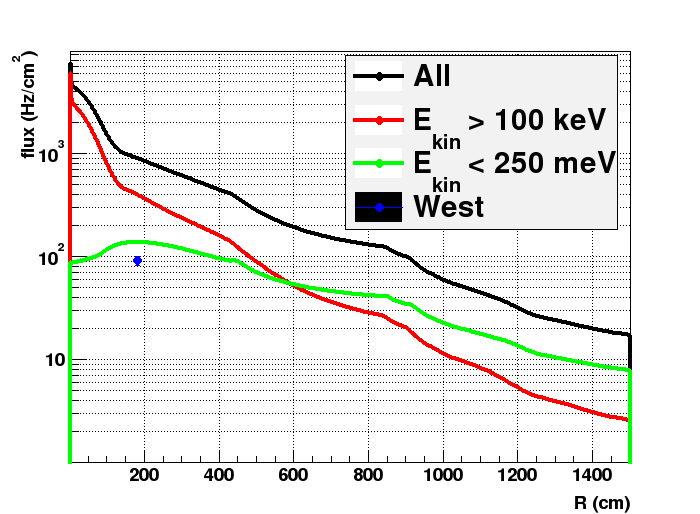} 
  } 
  \caption{The radial dependence of fluxes at Z = 0 (a) and Z = 675 cm (b) for all neutrons, neutrons with $E_{kin} > 100 keV$ and $E_{kin} < 250 meV$, and the measured flux at the \textbf{South}, \textbf{North} and  \textbf{West}  locations.}
  \label{fig:NeutronsZ}
\end{figure}

Flux was defined as sum of track length of a particle collected in a given volume in unit time divided by the volume size.
The fluxes were normalized  to  1 MHz rate of pp inelastic events at  $\sqrt s$ = 510 GeV.
Fluxes for all neutrons and  neutrons with $E_{kin} < 250 meV$ are show in Fig.\ref{fig:Neutron} and Fig.\ref{fig:ThermalNeutron}, respectively.
The radial dependence of fluxes at Z $\approx $ 0 and Z $\approx $ 675 cm 
for all neutrons, neutrons with $E_{kin} > 100 keV$ and neutrons with $E_{kin} < 250 meV$
are shown in Fig.\ref{fig:NeutronsZ}.

\section{Conclusions}
From this study we conclude that 
we can estimate neutron background for STAR detector with good precision. 
The results of the measurement and simulation are presented with absolute values and their ratios in Table \ref{table}.
The comparison is good (within $30\%$) for the \textbf{West}, \textbf{East} and \textbf{Far Away} locations. 
However, for the \textbf{South}, \textbf{North} and \textbf{Bottom} locations the simulation overestimated flux by a factor of $\approx$ 3. 
This conclusion is very close to one  \cite{ATLAS-MPX}  which we cited in the introduction.
The mismatch between the measurement and the simulation may be due to inaccurate description of geometry and material in the WAH, which would affect 
the neutron dissipation from the interaction region.
The deviation could also be related to the neutron transport parameters.

\section{Acknowledgments}
 We thank Brookhaven National Laboratory Instrumentation Division and, especially, G.Smith and N. Schaknowski, for the $He^3$ detector.
We thank the STAR Collaboration, the RHIC Operations Group and RCF at BNL. 
This work was supported by the Offices of NP and HEP within the U.S. DOE Office of Science.

\bibliographystyle{elsarticle-num}

\begin{thebibliography}{99}
\bibitem{STARcoord} STAR uses a right-handed coordinate system with its origin at the nominal interaction point and 
z-axis coinciding the the axis of the beam pipe. The x-axis points south,  the y-axis points upward, and the z-axis points to the west. STAR Note 0229A.
\end{thebibliography}

\end{document}